# COVID-19 Detection Using Deep Convolutional Neural Networks and Binary-Differential-Algorithm-Based Feature Selection From X-Ray Images

Mohammad Saber Iraji [a,c], Mohammad-Reza Feizi-Derakhshi [b, *], Jafar Tanha [c]

[a] Department of Computer Engineering and Information Technology, Payame Noor University, Tehran, Iran

[b] Computerized Intelligence Systems Laboratory, Department of Computer Engineering, University of Tabriz, Tabriz, Iran

[c] Department of Computer Engineering, University of Tabriz, Tabriz, Iran

Corresponding author: Mohammad-Reza Feizi-Derakhshi (mfeizi@tabrizu.ac.ir)

*Abstract*

The new Coronavirus is rapidly spreading and has already claimed the lives of numerous people. The virus is highly destructive to the human lungs, and early detection is critical. As a result, this paper presents a hybrid approach based on Deep Convolutional Neural Networks that are very effective tools for image classification. The feature vectors were extracted from the images using a deep convolutional neural network, and the binary differential metaheuristic algorithm was used to select the most valuable features. The SVM classifier was then given these optimized features. For the study, a database containing images from three categories, including COVID-19, pneumonia, and a healthy category, included 1092 X-ray samples, were used. The proposed method achieved a 99.43% accuracy, a 99.16% sensitivity, and a 99.57% specificity. Our findings indicate that the proposed method outperformed recent studies on COVID-19 detection using X-ray images.

1. INTRODUCTION

COVID-19's rapid spread has resulted in the death of numerous people worldwide. Muscle aches, cough, and fever are all symptoms of the virus, which can be detected through clinical trials and radiographic imaging.

Medical imaging is critical for disease diagnosis, and disease X-rays and computed tomography (CT) scans can be used in the deep network to aid in the disease's diagnosis.

The process of classifying and diagnosing disease from an image using a neural network is divided into four steps: feature extraction, optimal feature selection, network training, and model performance test. The feature extraction step is divided into two types. In the first type, image processing techniques, algorithms, and filters extract the features. Among the features extracted from the images are the tissue shapes and textures used to classify patients. In the second type, the original images and their actual output class are fed into the convolution network as input data, and the features are extracted automatically in the final flattened layer following the network training process and weight adjustment.

Certain features extracted from the deep network may have a detrimental effect on classification accuracy [1]. As a result, effective feature selection methods are critical. There are three distinct types of feature selection methods. The filter method uses features' intrinsic properties and statistical indicators such as the fisher score, information gain, chi-square, and correlation coefficient. The wrapper method employs a learning algorithm that searches the feature space for a subset of features that optimize the classification accuracy. To this end, wrapper approaches employ metaheuristic methods for selecting feature subsets and performing cross-validation. Finally, the hybrid method employs both filter and wrapper methods [2]. Metaheuristic methods outperform other feature selection methods in applications where many features are required.

Classification performance is improved by analyzing extracted features from images and selecting the optimal features [3]. Numerous feature selection (FS) studies have been published in the field of medical imaging, including Robustness-Driven FS (RDFS) for lung CT images [4], Shearlet transform FS from brain MRI images [5], principal component analysis for lung X-Ray images [6], genetic algorithm (GA) for lung nodules [7], bat algorithm (BA) versus particle swarm optimization (PSO) in lung X-ray images, and the flower pollination algorithm (FPA) from lung images [8].

The studies above propose that machine vision combined with metaheuristic algorithms can classify patients using lung images. On the other hand, existing diagnostic methods for the COVID-19 virus using X-ray images require a large amount of memory, ample time, and a large number of features. As a result, an intelligent system appears necessary to assist doctors and treatment staff in accurately and rapidly classifying COVID-19 patients in reducing disease-related mortality. This research aims to develop an efficient procedure utilizing artificial intelligence methods to assist doctors and patients in accurately predicting COVID-19. The research is novel in that it employs a binary differential evolution algorithm to design a deep learning structure based on feature selection for COVID-19 diagnosis. The contributions of the study include:

1- Using a deep convolutional neural network without a pre-trained network to design an intelligent system based on lung X-ray images and extracting features with the least amount of memory required to create and train the network.

2- Selecting the optimal features of the differential metaheuristic method that improves performance indexes.

3- Increasing classification accuracy for multi-class problems, including patients with COVID-19, pneumonia, and the healthy group.

The study is structured as follows: Section 2 reviews related works. Section 3 presents the proposed methodology and model for COVID-19 detection using deep convolution and binary differential algorithms. Section 4 contains the experimental results, and section 5 discusses the method and compares this with prior works. Finally, the study concludes.

**2. RELATED WORKS**

Hemdan, Shouman, and Karar used deep learning models to infer the positive or negative status of COVID and reported that the VGG19 model performed better with an accuracy of 90% on 25 COVID infected and 25 non-COVID images [9]. Toaçar, Ergen, and Cömert incorporated 295 COVID images, 98 pneumonia images, and 65 normal images into MobileNet and SqueezeNet [10]. They extracted features from trained Net models and then used the SMO algorithm to select the features, with an overall accuracy of 99.27% reported for the SVM classifier. Zhang, Xie, Li, Shen, and Xia investigated an 18-layer ResNet model for 100 COVID and 1431 pneumonia images and reported an accuracy of 95.18% [11]. Apostolopoulos & Mpesiana pre-trained VGG19 on 224 COVID, 700 pneumonia, and 504 normal images, where the results demonstrated a 98.75% accuracy [12]. The authors of [13] evaluated the DarkNet with 17 convolutional layers using 127 COVID, 500 pneumonia, and 500 normal images and reported an accuracy of 98.08%. In [14], the performance of CNN was improved via preprocessing image algorithms, resulting in a model with 94.5% accuracy.

The authors of [1] developed a COVID-19 classification method based on two datasets that combined a CNN named Inception, a pre-trained Imagnet as a feature extractor, the Marine Predators Algorithm as a feature selector, and a KNN as a classifier.

The first dataset contained 200 positive COVID-19 images and 1675 negative images, whereas the second dataset contained 219 positive COVID-19 images and 1341 negative images. Accuracy was reported as 98.7% for dataset 1 and 99.6% for dataset 2. Canayaz validated a COVID-19 diagnosis model by combining VGG19,

ResNet, AlexNet, and GoogLeNet with two metaheuristic algorithms titled "binary particle swarm optimization" and "binary gray wolf optimization." The highest overall classification accuracy was 99.38% after binary gray wolf optimization was used to select features from 1092 X-ray images from the COVID-19, pneumonia, and healthy category records [15].

One of the previous works' limitations is their reliance on pre-trained deep networks, which require a large amount of memory. Additionally, many input features plus a lengthy detection time are further drawbacks of these methods. In this study, a deep learning approach based on feature selection is proposed via the binary differential evolution algorithm to overcome these limitations and improve COVID-19 detection.

## 3. METHODOLOGY AND MODEL

Figure 1 depicts the proposed model. Initially, the convolutional neural network is fed with lung images. After training the network, features are extracted from suboptimal images. The heuristic method is then used to extract the optimal features. As a result, the three classes of COVID-19, pneumonia and healthy, are classified with higher accuracy.

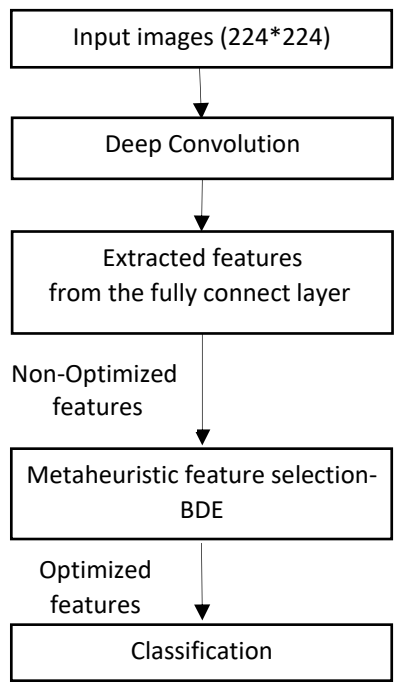

Figure 1. The proposed model for COVID-19

*3.1. Deep convolution*

Convolutional neural networks are used in machine learning as a feature extractor and classification method. The input to a convolutional network is the original data, such as images. The network extracts the features automatically using the convolution function. After learning, rather than manually extracting the feature, the matrixes serve as filters that slide over the main input image, and the convolution operation is carried out via Equation 1. Finally, after training and mapping the input images to the output labels, several convolution layers extract the features [16].

$$(IMG * C)_{ij} = \sum_{p=0}^{c1-1} \sum_{q=0}^{c2-1} \sum_{c=1}^{tc} C_{p,q,c} \cdot IMG_{i+p,j+q,c} + bs \qquad (1)$$

where IMG denotes the input image with height =H, width=W dimensions, and tc is the number of image channels, C is the filter matrix with c1*c2 dimensions, and bs is a bias value for each filter C, i=0…H, j=0…W

Following convolution, the unwanted values are removed using the ReLu layer, and the input is then reduced using the pooling layer. The effective input vector is then passed to the fully connected layer, which functions similarly to the MLP. In the final section of the deep convolution layers, Softmax [17], classification layers perform classification using ADAM (adaptive moment optimizer) [18], the lost function is shown below (equation 2).

$$L(w,b) = -\frac{1}{M}\sum_{m=1}^{M}[y_m \log \hat{y}_m + (1-y_m)\log(1-\hat{y}_m)] + \Gamma \times \sum_{r=1}^{M}\|w^r\|_2 \qquad (2)$$

where M denotes the sample size, $y_m$ denotes the actual class for the mth sample, $\hat{y}_m$ denotes the predicted output class for the mth input data, and Γ denotes the regularization coefficient.

ADAM is a gradient-based optimization algorithm that uses the exponential moving average of the gradient and the square of the gradient to update the neural network weights and solve deep network issues effectively. The deep neural network comprises numerous layers, each with its own set of learning parameters, namely weights and biases. Applying the optimal feature selection algorithm to the ADAM optimizer increases the optimization's speed and accuracy.

*3.2. Binary Differential Evolution*

Differential evolution (DE) [19] is a heuristic evolutionary method for minimizing the continuous problem. The concept of binary differential evolution (BDE) [20] is extended to address issues of feature selection. It is composed of three distinct builders, including mutation, crossover, and selection. Initially, dimensions D are used to generate the initial population, where D is the number of features to optimize. For the mutation operation, three random vectors $p_{u1}, p_{u2}, p_{u3}$ are selected for vector $p_k$ such that $u1 \neq u2 \neq u3 \neq k$., k is a population vector arrangement.

If the dth dimensions of vectors $p_{u1}$ and $p_{u2}$ are equal, the dth feature of the difference vector (Equation 3) is zero; otherwise, it has the same value as vector $p_{u1}$.

$$difference\ vector_k^d = \begin{cases} 0 & p_{u1}^d = p_{u2}^d \\ p_{u1} & other \end{cases} \quad (3)$$

Afterward, the mutation and crossover operations are executed, as shown in Equations 4 and 5.

$$mute\ vector_k^d = \begin{cases} 1, & if\ difference\ vector_k^d = 1 \\ p_{u3}^d, & other \end{cases} \quad (4)$$

$$W_k^d = \begin{cases} mute\ vector_k^d, & if\ \gamma \leq CR \parallel d = d_{random} \\ p_k^d, & other \end{cases} \quad (5)$$

where W represents the try vector, CR $\epsilon$ (0, 1) represents the crossover amount, and $\gamma\ \epsilon$ (0, 1) represents a random number. In the selection procedure, if the fitness value of the try vector $W_k$ is greater than that of the current vector $p_k$, it will be replaced. Otherwise, the current vector $p_k$ is stored for the next generation.

## 4. EXPERIMENTAL RESULTS

### 4.1. Description of data

Canayaz developed a COVID-19 X-ray data set that included three subgroups of patients, including those with COVID-19, those with pneumonia, and those who were healthy [15]. By combining data from this database, a total of 364 images for each of the three categories were obtained as a balanced dataset [21-23]. The total number of images is equal to the number of classes multiplied by the number of class instances = (3*394)1092, with a 224 x 224 dimension. The same data is used in this study to predict COVID-19 disease using a convolutional neural network and to select optimal features using the binary differential metaheuristic algorithm. Figure 2 illustrates a representative sample of three output classifications: COVID-19, pneumonia, and healthy.

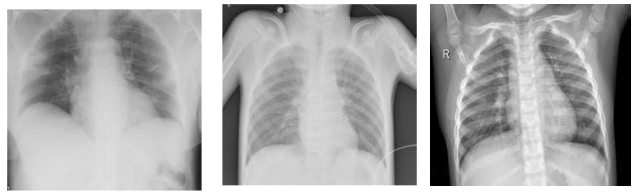

Figure 2. Chest X-ray images of the different conditions (a) COVID-19 (b) Pneumonia (c) Healthy

### 4.2. Performance evaluation

The proposed model was run in MATLAB version 9.1.0.441655 (R2018b) on a laptop computer equipped with a 1.8 GHz processor and 4 Gigabytes of RAM. After training in the application phase, the proposed method took an average of 29 seconds per patient, which can be reduced by improving the hardware technology used. The COVID-19 prediction model was evaluated using the accuracy, sensitivity, specificity, geometric mean, and area under the curve (AUC-ROC) [24, 25] performance metrics (Equations 6-9), where accuracy refers to the correctness of the classification. The proportion of correctly distinguished negative cases is referred to as "specificity," while the proportion of correctly distinguished positive cases is referred to as "sensitivity." The geometric mean is the second root of the sensitivity and specificity products. Higher values of the area under the curve (AUC) within the receiver operating characteristics indicate improved classification performance.

$$Accuracy = \frac{TP+TN}{TP+TN+FP+FN} \quad (6)$$

where:

True positives = the number of samples that are correctly labeled as positive

False positives = the number of samples that are wrongly labeled as positive

True negatives = the number of samples that are correctly labeled as negative

False negatives = the number of samples that are wrongly labeled as negative

Sensitivity = TP / (TP+ FN)     (7)

Specificity = TN / (FP+TN)     (8)

Geometric mean = $\sqrt[2]{Sensitivity * Specificity}$     (9)

*4.3. Model parameters*

Figure 3 illustrates the network structure of a deep convolutional neural network. Firstly, the image input layer dimensions were 224*224, and the convolution operator used eight 3-by-3 filters. After processing the first block of the network layers, i.e., image input, convolution, Batch Normalize, ReLU, max-pooling layers, fully connected layer 400, ReLU, Drop out, the local features were automatically extracted. Finally, the second network block categorized the input images into three output classes by utilizing three fully connected layers, softmax, and classification. After 200 epochs, the validation accuracy was 97.25% when using the ADAM optimizer, and the mini-batch size was 64 (Figure 4). Due to the neural network's regularization and barricade overfitting, batch normalization and dropout were used.

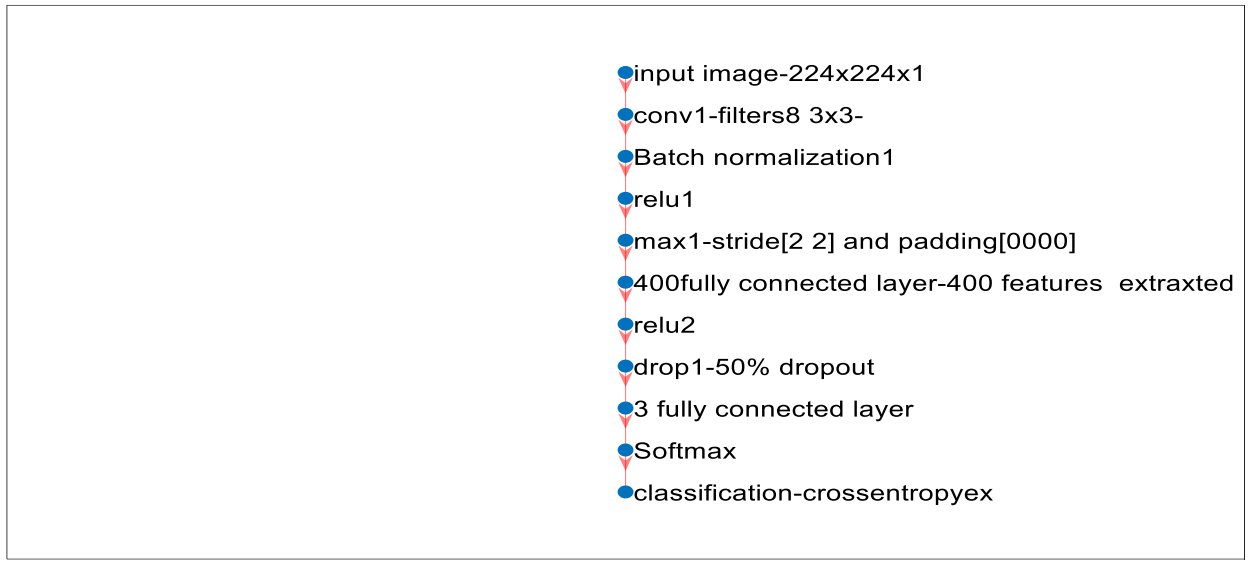

Figure. 3. The proposed model layers' structure

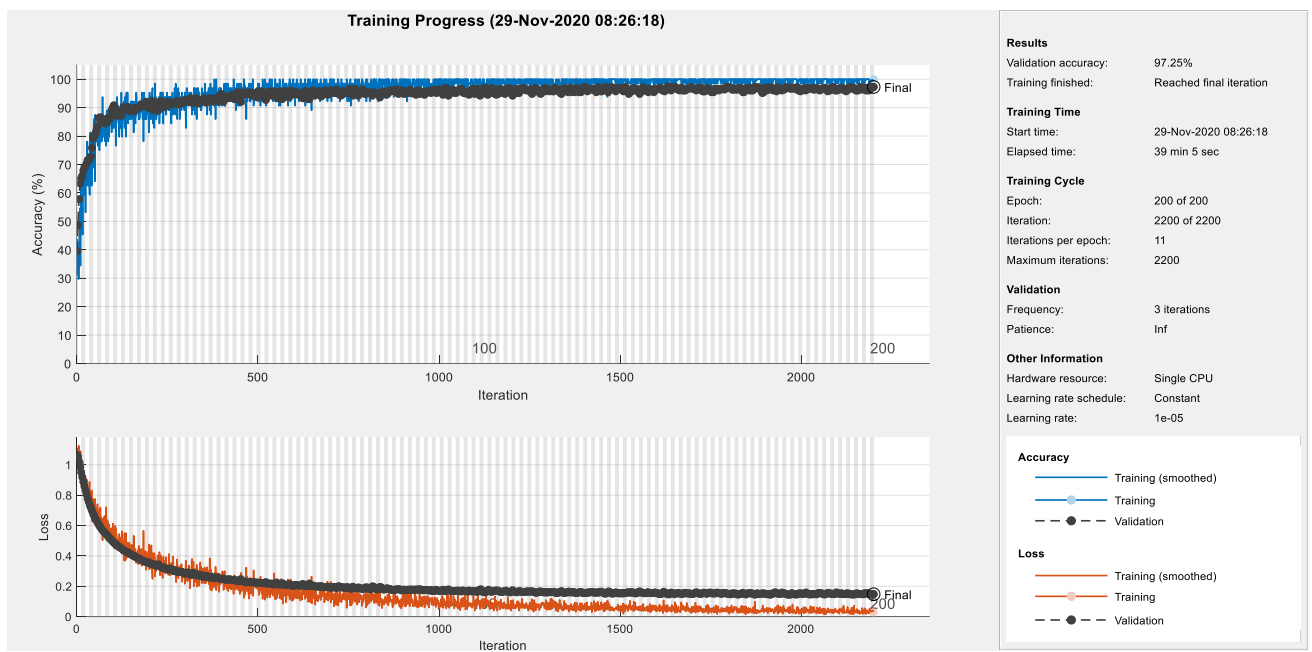

Figure. 4. Accuracy and loss measurements for convolutional neural networks used for training and validation data

Convolutional networks are used to transform data into feature vectors. Given that some network features may degrade the model's performance [1], after extracting 400 features in the first fully connected layer, the binary differential metaheuristic algorithm was used to select the optimal feature subset and eliminate unnecessary features. The binary differential algorithm's parameters were population = 20, iteration = 100 (figure 5), and a crossover rate of 1. The amount (1- (geometric mean)) of the SVM classifier [26] was regarded as the population's fitness values (Figure 5). Following the binary differential algorithm's execution, 340 optimal features were selected.

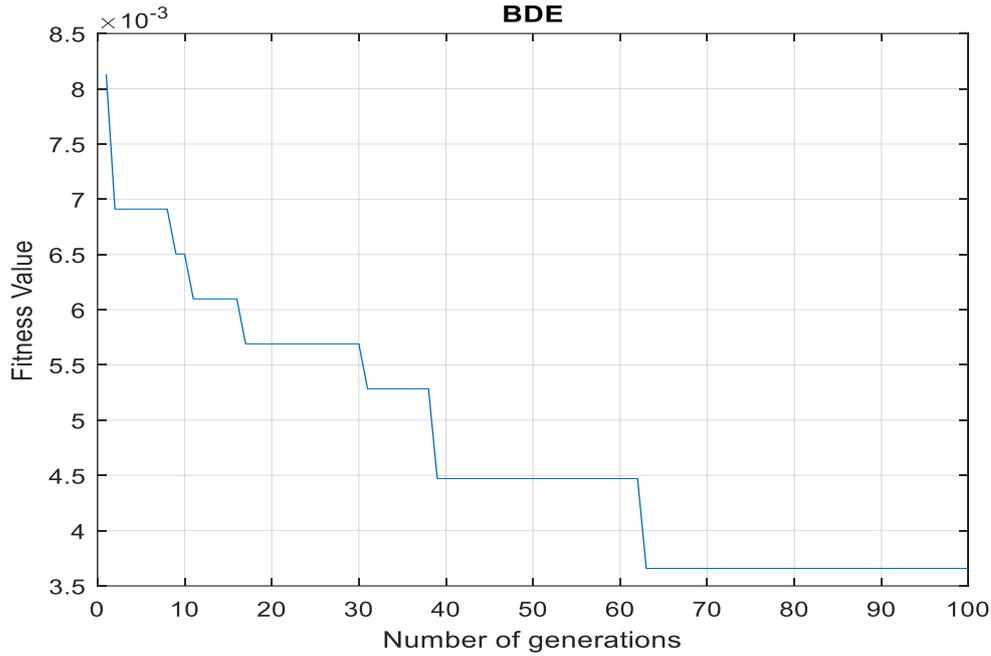
Figure. 5. Fitness curve for the binary differential algorithm

*4.4. Performance comparison*

The conventional validation (CV) method, which employs random sampling, is one of the training and testing protocols used to determine the model's accuracy and validate the estimation results. According to the CV method, 70% of data was used for training, 15% for validation, and 15% for testing [27, 28]. The proposed method was applied to the data, and 100 runs were performed to prevent overfitting [29]. The optimally selected features from the differential algorithm and the initial extracted features from the deep convolutional neural network were entered into the SVM classifier.

The confusion matrix for the SVM classifier's original and optimized features is demonstrated in Table 1 using training, testing, validation, and total data.

Table 1 depicts the confusion matrix using training, testing, validation, and total data for the SVM classifier's original features and optimized features. The TP, TN, FP, FN, accuracy, sensitivity, specificity, geometric mean, AUC metrics for each of the three output classes plus each type of testing, training, validation, and total data were computed (Table 2).

## 5. DISCUSSION

Predicting a disease can be accomplished by combining images with a deep neural network, where a deep neural network can be used as a feature extractor. The large size and volume of images applied to the deep neural network results in numerous feature formations that increase the training and decision times of the predictive model.

The proposed model faced several design challenges, including collecting and improving lung images, the deep network architecture, in terms of the structure, number, and type of layers, plus the metaheuristic algorithm, the initial population, and the metaheuristic algorithm's objective function type. The presence of inefficient features extracted from the deep network may reduce the predictive model's accuracy and efficiency; thus, using the metaheuristic method to select the optimal features improved the model's memory, time, and accuracy.

According to Table 3, the proposed model achieved an accuracy of 99.43%, a sensitivity of 99.16%, a specificity of 99.57%, a gmean of 99.37%, an AUC of 0.99, and a root mean square error (RMSE) of 0.1133 using features extracted from the X-ray image via the CNN and features optimized using the binary differential metaheuristic algorithm. The accuracy of the classification of the COVID-19 problem was calculated to be 99.43% in this study, and the number of relevant features was 304 (Table 4), whereas, in a previous study [15], these figures were reported to be 99.38% and 448 features, respectively, based on the same data.

Transfer learning models are trained to classify 1,000 different types of object images and must be retrained to classify specific issues such as COVID-19 detection. Although the learning process is prompt in models like ResNet and SqueezeNet, they require preprocessing the input image, sizing the data set, and setting multiple parameters. The upper layers extract color and edge features, while the deeper layers extract complex features. Process time increases as the number of layers in transfer learning models increases. The trained transfer learning model's feature map and activation layers must be customized for the specific COVID-19 problem, which requires a large amount of memory. After fine-tuning the pre-trained model's principal component analysis (PCA), the optimal feature can be selected using heuristic methods, automated encoders, or variance-based selectors. Finally, ensemble methods, such as a combination of SVMs or other classifiers, can be used to predict COVID-19 disease diagnosis accuracy. Using semi-supervised self-learning methods may result in acceptable accuracy and reduced labeling time.

In future work, a different feature selection algorithm and the application of additional learners may produce improved results. Along with the images, the parameters derived from clinical trials can create a new model with a novel combination of features for diagnosing the disease and possibly predicting mortality as a result.

Table 1. Confusion matrix average after 100 runs with 3 classes using the (a) optimized features (b) original features based on the training, test, validation, and total data

(a) Optimized features

| Optimized features test | | Predicted | | |
|---|---|---|---|---|
| | | Covid | Normal | Pneumonia |
| Actual | Covid | 51/75 | 0 | 0/2 |
| | Normal | 0/15 | 54/9 | 0/1 |
| | Pneumonia | 0/4 | 0/55 | 55/95 |

| Optimized features validation | | Predicted | | |
|---|---|---|---|---|
| | | Covid | Normal | Pneumonia |
| Actual | Covid | 55/05 | 0/05 | 0/45 |
| | Normal | 0/05 | 54/4 | 0/3 |
| | Pneumonia | 0/4 | 1 | 52/3 |

| Optimized features training | | Predicted | | |
|---|---|---|---|---|
| | | Covid | Normal | Pneumonia |
| Actual | Covid | 256/5 | 0 | 0 |
| | Normal | 0 | 254/1 | 0 |
| | Pneumonia | 0 | 0 | 253/4 |

| Optimized features total | | Predicted | | |
|---|---|---|---|---|
| | | Covid | Normal | Pneumonia |
| Actual | Covid | 363/3 | 0/05 | 0/65 |
| | Normal | 0/2 | 363/4 | 0/4 |
| | Pneumonia | 0/8 | 1/55 | 361/65 |

(b) Original features

| Original features test | | Predicted | | |
|---|---|---|---|---|
| | | Covid | Normal | Pneumonia |
| Actual | Covid | 51/65 | 0 | 0/3 |
| | Normal | 0/25 | 54/65 | 0/25 |
| | Pneumonia | 0/5 | 0/95 | 55/45 |

| Original features validation | | Predicted | | |
|---|---|---|---|---|
| | | Covid | Normal | Pneumonia |
| Actual | Covid | 54/85 | 0 | 0/7 |
| | Normal | 0/15 | 54/2 | 0/4 |
| | Pneumonia | 0/55 | 1/2 | 51/95 |

| Original features training | | Predicted | | |
|---|---|---|---|---|
| | | Covid | Normal | Pneumonia |
| Actual | Covid | 256/45 | 0 | 0/05 |
| | Normal | 0 | 254/1 | 0 |
| | Pneumonia | 0 | 0 | 253/4 |

| original features total | | Predicted | | |
|---|---|---|---|---|
| | | Covid | Normal | Pneumonia |
| Actual | Covid | 362/95 | 0 | 1/05 |
| | Normal | 0/4 | 362/95 | 0/65 |
| | Pneumonia | 1/05 | 2/15 | 360/8 |

Table 2. Comparison of indicators (TP, TN, FP, FN, accuracy, the area under curve, sensitivity, specificity, geometric mean) for any output class based on (a) optimized features (b) original features

(a) Optimized features

| Optimized features –total | Covid | Normal | Pneumonia |
|---|---|---|---|
| TP | 363/30 | 363/40 | 361/65 |
| TN | 727/00 | 726/40 | 726/95 |
| FP | 1/00 | 1/60 | 1/05 |
| FN | 0/70 | 0/60 | 2/35 |
| Accuracy | 99/84 | 99/80 | 99/69 |
| Sensitivity | 99/81 | 99/84 | 99/35 |
| Specificity | 99/86 | 99/78 | 99/86 |
| Geometric mean | 99/84 | 99/81 | 99/60 |
| Area under curve | 0/9984 | 0/9981 | 0/9961 |

| Optimized features –training | Covid | Normal | Pneumonia |
|---|---|---|---|
| TP | 256/50 | 254/10 | 253/40 |
| TN | 507/50 | 509/90 | 510/60 |
| FP | 0/00 | 0/00 | 0/00 |
| FN | 0/00 | 0/00 | 0/00 |
| Accuracy | 100/00 | 100/00 | 100/00 |
| Sensitivity | 100/00 | 100/00 | 100/00 |
| Specificity | 100/00 | 100/00 | 100/00 |
| Geometric mean | 100/00 | 100/00 | 100/00 |
| Area under curve | 1/0000 | 1/0000 | 1/0000 |

| Optimized features –test | Covid | Normal | Pneumonia |
|---|---|---|---|
| TP | 51/75 | 54/90 | 55/95 |
| TN | 111/50 | 108/30 | 106/80 |
| FP | 0/55 | 0/55 | 0/30 |
| FN | 0/20 | 0/25 | 0/95 |
| Accuracy | **99/54** | **99/51** | **99/24** |
| Sensitivity | 99/62 | 99/55 | 98/33 |
| Specificity | 99/51 | 99/49 | 99/72 |
| Geometric mean | 99/56 | 99/52 | 99/02 |
| Area under curve | 0/9956 | 0/9952 | 0/9903 |

| Optimized features -valid | Covid | Normal | Pneumonia |
|---|---|---|---|
| TP | 55/05 | 54/40 | 52/30 |
| TN | 108/00 | 108/20 | 109/55 |
| FP | 0/45 | 1/05 | 0/75 |
| FN | 0/50 | 0/35 | 1/40 |
| Accuracy | 99/42 | 99/15 | 98/69 |
| Sensitivity | 99/10 | 99/36 | 97/39 |
| Specificity | 99/59 | 99/04 | 99/32 |
| Geometric mean | 99/34 | 99/20 | 98/35 |
| Area under curve | 0/9934 | 0/9920 | 0/9840 |

(b) Original features

| Original features -total | Covid | Normal | Pneumonia |
|---|---|---|---|
| TP | 362/95 | 362/95 | 360/80 |
| TN | 726/55 | 725/85 | 726/30 |
| FP | 1/45 | 2/15 | 1/70 |
| FN | 1/05 | 1/05 | 3/20 |
| Accuracy | 99/77 | 99/71 | 99/55 |
| Sensitivity | 99/71 | 99/71 | 99/12 |
| Specificity | 99/80 | 99/70 | 99/77 |
| Geometric mean | 99/76 | 99/71 | 99/44 |
| Area under curve | 0/9976 | 0/9971 | 0/9944 |

| Original features -training | Covid | Normal | Pneumonia |
|---|---|---|---|
| TP | 256/45 | 254/10 | 253/40 |
| TN | 507/50 | 509/90 | 510/55 |
| FP | 0/00 | 0/00 | 0/05 |
| FN | 0/05 | 0/00 | 0/00 |
| Accuracy | 99/99 | 100/00 | 99/99 |
| Sensitivity | 99/98 | 100/00 | 100/00 |
| Specificity | 100/00 | 100/00 | 99/99 |
| Geometric mean | 99/99 | 100/00 | 100/00 |
| Area under curve | 0/9999 | 1/0000 | 1/0000 |

| Original features -test | Covid | Normal | Pneumonia |
|---|---|---|---|
| TP | 51/65 | 54/65 | 55/45 |
| TN | 111/30 | 107/90 | 106/55 |
| FP | 0/75 | 0/95 | 0/55 |
| FN | 0/30 | 0/50 | 1/45 |
| Accuracy | 99/36 | 99/12 | 98/78 |
| Sensitivity | 99/42 | 99/09 | 97/45 |
| Specificity | 99/33 | 99/13 | 99/49 |
| Geometric mean | 99/38 | 99/11 | 98/46 |
| Area under curve | 0/9937 | 0/9910 | 0/9847 |

| Original features -valid | Covid | Normal | Pneumonia |
|---|---|---|---|
| TP | 54/85 | 54/20 | 51/95 |
| TN | 107/75 | 108/05 | 109/20 |
| FP | 0/70 | 1/20 | 1/10 |
| FN | 0/70 | 0/55 | 1/75 |
| Accuracy | 99/15 | 98/93 | 98/26 |
| Sensitivity | 98/74 | 99/00 | 96/74 |
| Specificity | 99/35 | 98/90 | 99/00 |
| Geometric mean | 99/05 | 98/95 | 97/87 |
| Area under curve | 0/9904 | 0/9895 | 0/9789 |

Table 3. Average of the confusion matrix components after 100 runs using original and optimized features

| Method | TP | TN | FP | FN | Accuracy | Sensitivity | Specificity | Geometric mean | Area Under Curve | RMSE |
|---|---|---|---|---|---|---|---|---|---|---|
| Original features via Deep convolution | | | | | | | | | | |
| Training | 254/65 | 509/32 | 0/02 | 0/02 | 1/0000 | 0/9999 | 1/0000 | 1/0000 | 0/9999 | 0/0036 |
| Testing | 53/92 | 108/58 | 0/75 | 0/75 | 0/9909 | 0/9866 | 0/9931 | 0/9898 | 0/9898 | 0/1533 |
| Validation | 53/67 | 108/33 | 1/00 | 1/00 | 0/9878 | 0/9816 | 0/9909 | 0/9862 | 0/9863 | 0/1905 |
| Total | 362/23 | 726/23 | 1/77 | 1/77 | 0/9968 | 0/9951 | 0/9976 | 0/9964 | 0/9964 | 1/1543 |
| Optimized features via Binary Deferential | | | | | | | | | | |
| Training | 254/67 | 509/33 | 0/00 | 0/00 | 1/0000 | 1/0000 | 1/0000 | 1/0000 | 1/0000 | 0/0000 |
| Testing | 54/20 | 108/87 | 0/47 | 0/47 | 0/9943 | 0/9916 | 0/9957 | 0/9937 | 0/9937 | 0/1133 |
| Validation | 53/92 | 108/58 | 0/75 | 0/75 | 0/9909 | 0/9862 | 0/9931 | 0/9896 | 0/9898 | 0/1592 |
| Total | 362/78 | 726/78 | 1/22 | 1/22 | 0/9978 | 0/9967 | 0/9983 | 0/9975 | 0/9975 | 1/1543 |

Table 4. A comparison of the proposed method with prior research

| Research | Method | Number of features | Accuracy | Geometric mean | RAM | Max Computation time |
|---|---|---|---|---|---|---|
| [15] | Binary particle swarm optimization -VGG19 | 448 | 99.38 | - | 16 gigabyte | 2500s |
| Proposed method | Binary differential -cnn | 308 | 99.43 | 99.37 | 4 gigabyte | 2300s |

## 6. CONCLUSION

The number of people infected with COVID-19 has risen rapidly. Machine vision techniques and artificial intelligence are critical in diagnosing and treating disease. The purpose of this paper was to propose a method for the "COVID-19" problem via a set of lung images that included three categories of pneumonia, COVID-19, and healthy.

A deep convolutional neural network consisting of 11 layers was applied to extract the features. The binary differential metaheuristic method was used to select relevant features and eliminate unrelated features. Lung X-ray images were classified using an SVM classifier based on these optimal features. This study demonstrated that the accuracy indicator and the number of relevant extracted features outperformed previous methods using the same data. Based on a deep neural network and a metaheuristic feature selection algorithm, the proposed model can be used in various other medical applications.

Acknowledgments

The authors wish to express their gratitude to all study participants.

Authors' contributions

Iraji and Feizi-Derakhshi proposed the image analysis algorithm; Iraji implemented the algorithm and analyzed the experimental results; Tanha provided clinical guidance; and Iraji, Feizi-Derakhshi, and Tanha validated the obtained results. The final manuscript was read and approved by all authors.

Funding

This research received no specific funding from any public, commercial, or not-for-profit funding agency.

Availability of data and materials

The datasets used and analyzed in this study are available upon reasonable request from the corresponding author.

DISCLOSURE OF POTENTIAL CONFLICTS OF INTEREST

Conflict of interest

The authors declare no potential conflicts of interest related to the research, authorship, or publication of this article.

Ethical approval

This paper contains no data or other information derived from studies or experiments involving human or animal subjects.

Consent for publication

Agreed by the authors.

Competing interests

The authors declare that no competing interests exist.